\documentclass[%
 preprint,
 aps,
 nofootinbib,
]{revtex4-2}

\usepackage{mathrsfs,mathpazo,amsmath,mathtools,physics}

\usepackage{graphicx}
\usepackage{dcolumn}
\usepackage{url}
\usepackage{hyperref}
\usepackage{xcolor}
\usepackage{ulem}
\usepackage{tikz}
\usepackage{listings}
\usepackage{lipsum, babel}
\usepackage{multirow}
\usepackage[font=small,labelfont=bf,justification=justified]{subcaption}
\usepackage[font=small,labelfont=bf,justification=justified]{caption}

\usepackage[makeroom]{cancel}
\usepackage{float}
\usepackage{pgf}

\tikzset{
    level/.style = {
        ultra thick,
        black,
    },
    connect/.style = {
        dashed,
        red
    }
}
\usetikzlibrary{backgrounds,calc}
\usetikzlibrary{shapes.misc}
\usepackage{multirow}
\allowdisplaybreaks
\makeatletter
\newcommand*\rel@kern[1]{\kern#1\dimexpr\macc@kerna}
\newcommand*\widebar[1]{%
  \begingroup
  \def\mathaccent##1##2{%
    \rel@kern{0.8}%
    \overline{\rel@kern{-0.8}\macc@nucleus\rel@kern{0.2}}%
    \rel@kern{-0.2}%
  }%
  \macc@depth\@ne
  \let\math@bgroup\@empty \let\math@egroup\macc@set@skewchar
  \mathsurround\z@ \frozen@everymath{\mathgroup\macc@group\relax}%
  \macc@set@skewchar\relax
  \let\mathaccentV\macc@nested@a
  \macc@nested@a\relax111{#1}%
  \endgroup
}

\newcommand{\der}[2]{\frac{\partial #1}{\partial #2}}

\newcommand{\dder}[2]{\frac{\partial^2 #1}{\partial #2 ^2}}
\let\ph\varphi
\let\tet\theta
\let\la\lambda
\begin{document}

\title{Magnetized neutron stars: perturbative versus fully-numerical
  approaches}

\author{Debarati Chatterjee}
\email{debarati@iucaa.in}
\affiliation{Inter-University Centre for Astronomy and Astrophysics, Puna, India}
\author{Daw Guttmann}
\affiliation{Observatoire de Paris, Universit\'e PSL, Sorbonne Universit\'e, CNRS, UMR 7095, Institut d'Astrophysique de Paris, 75014 Paris, France}
\author{J\'er\^ome Novak}
\email{jerome.novak@astro.unistra.fr}
\affiliation{Observatoire astronomique de Strasbourg, CNRS, Université de Strasbourg, 11 rue de l'Université, 67000 Strasbourg, France}
\author{Micaela Oertel}
\email{micaela.oertel@astro.unistra.fr}
\affiliation{Observatoire astronomique de Strasbourg, CNRS, Université de Strasbourg, 11 rue de l'Université, 67000 Strasbourg, France}

\author{Martin Jakob Steil}
\affiliation{Institut f\"ur Kernphysik, Theoriezentrum,	Technische Universit\"at Darmstadt,	D-64289 Darmstadt, Germany}

\date{\today}

\begin{abstract}
  \begin{description}
  \item[Background] for the study of highly magnetized neutron
  stars observed as magnetars, and to quantify the effect of this
  intense magnetic field onto the star's structure and shape which
  can be particularly relevant for the study of emission of continuous
  gravitational waves, both numerical and perturbative approaches have been developped.
  \item[Methods] we compare these two approaches in General
  Relativity with the limitation to the case where the magnetic field has a
  purely poloidal structure. The perturbative one~\cite{konno-99}
  assumes that the deformation induced by the magnetic field is small
  and that this field arises only from dipole currents. The full
  numerical one is based on the library
  \textsc{lorene}~\cite{lorene}. \item[Results] we have used both
  approaches to compute the magnetic field distribution and the
  deformation of the star, varying the value of the magnetic field at
  the pole, the compactness of the star and its equation of state. (4)
  Conclusions: whereas the perturbative approach breaks down for very
  high polar magnetic field values (typically above a few times
  $10^{16}$~G), it gives very good results for observed values, even
  in magnetars. On the contrary, the numerical code exhibits
  resolution problems for relatively low magnetic field values
  (typically $10^{10}$~G 
  ), which translates into imprecise computation of the star's
  deformation and mass quadrupole moment.
  \end{description}
  \end{abstract}

\maketitle
\section{Introduction\label{s:intro}}
Neutron stars represent perfect probes for testing our knowledge of
physics in extreme conditions. They are the most compact stars known,
being thus correctly described only within the theory of General
Relativity. Densities in their interior can reach values above
$10^{14}\textrm{ g cm}^{-3}$, that is several times the nuclear
saturation density and matter is very neutron rich in large parts of
the star. In this regime, it is not feasible to complete \textit{ab
  initio} calculations using quantum chromodynamics and matter is
typically described by effective models, which induces large
uncertainties about its composition and properties (see
e.g.~Ref.~\cite{oertel-17}). Magnetic field values deduced from the
measured pulsar spin-down rates can be as high as
$10^{15}$~G~\cite{olausen-14} in the case of
magnetars~\cite{rea-26}. However, these estimates rely on the
\textsl{vacuum rotating dipole model} for the pulsar emission and
magnetic field surface intensity could be different, if some more
elaborated models are taken into account~\cite{petri-19}. Still,
direct observations of cyclotron lines are also pointing toward
magnetic fields up to about $3\times
10^{14}$~G~\cite{ibrahim-04}. From another perspective, theoretical
estimates using the virial theorem give a maximal expected value of
about $10^{18}$~G at the star's center.

The question then arises about how much such strong magnetic fields
can influence the neutron star's structure. This is of particular
importance because such deformed and rotating neutron stars can be
sources of gravitational waves~\cite{bonazzola-96}.  Magnetic
deformations depend not only on the strength and configuration of the
magnetic field inside the star, but also on its interior composition
and equation of state (EoS), and therefore highly uncertain, given our
limited theoretical understanding of the structure and field topology
of strongly magnetised neutron stars.  Currently extensive searches
are made on the LIGO-Virgo-Kagra data to find so-called
\textsl{continuous gravitational waves} coming from known (but not
exclusively)
pulsars~\cite{lvk_pulsars-25,lvk_narrowband,lvk_allsky_isolated}.
In particular, searches are conducted for long-duration transients~\cite{lvk_allsky_transients}
 that target newly born neutron stars or magnetars resulting from core collapse 
of a massive star or mergers of binary neutron stars.
Detection of signals from such sources would put constraints on neutron star structure, 
EoS for dense matter (~\cite{dallosso-15,margalit-19}), magnetic field configuration 
(~\cite{mastrano-11,mastrano-15}), internal superconductivity (~\cite{das-26}), and tests of general relativity.  Modeling deformations in strongly magnetised neutron stars is a challenging problem, and several recent
works (see e.g.~\cite{pagliaro-25}) have highlighted the need for more
systematic theoretical investigations crucial for designing searches
for gravitational waves as well as correct astrophysical
interpretation from the derived upper limits following
detections. Current theoretical estimates
(~\cite{ushomirsky-00,horowitzkadau-09,moraleshorowitz-22,gittins-21})
indicate that continuous gravitational waves from magnetically
deformed neutron stars have small spin-down limit ellipticities and
may not be detectable by the current generation of ground-based
detectors, but could be potentially interesting targets in the deci-Hz
band such as DECIGO, Big Bang Observer or other proposed missions in
this frequency range, or by third generation proposed detectors such
as Cosmic Explorer and Einstein Telescope.  Therefore, it is quite
important to construct consistent theoretical models relating the
microphysics to global properties of neutron stars in presence of a
magnetic field.  Those microphyics models can, in turn, be influenced
by the magnetic field (for a recent review,
see~Ref.~\cite{adhikari-26} and references therein) and the
computation of global magnetic field distribution is thus also
relevant for dense matter models. Various approaches are possible
here, with different assumptions and ansatz, going from semi-analytic
models based on the perturbation of a spherically symmetric background
star~\cite{konno-99}, to full numerical ones following first studies
by e.g.~\cite{bocquet-95, cardall-01, kiuchi-08, ciolfi-10, ciolfi-13}.

In this work, we do not devise any new tool to study these
systems, but we compare the accuracy of both types of approaches, in
the specific case of non-rotating stars, with a poloidal magnetic
field and fulfilling axial symmetry. We want in particular to study
the respective validity of each type of models when computing the
magnetic field distribution and the deformation of the star, in view
of the analysis of gravitational wave emission. Please note that, from
the absence of magnetic monopoles, any physically relevant model must
depart from spherical symmetry and needs to be more involved. One
question is thus whether a full numerical approach, which often is
more difficult to handle, is always necessary to get a consistent model
of a magnetized neutron star, or whether a simpler, perturbative,
approach is sufficient. The paper is organized as
follows. Sec.~\ref{s:pert} presents the perturbative approach,
starting from the spherically symmetric (non-magnetized)
solution~\ref{ss:background}, followed by the perturbation equations
for the magnetic field~\ref{ss:pert_mag} and those for the deformation
of the star~\ref{ss:deform}. We then turn to the description of the
numerical approach in Sec.~\ref{s:magstar}, discussing first
in~\ref{ss:axi} some definitions linked to the general-relativistic
approach, including global quantities used for the comparison. The
central equations solved with this approach in the present work are 
presented in~\ref{ss:num_eqs}, before briefly sketching the numerical 
methods and tests~\ref{ss:num_tests} used. Sec.~\ref{s:compare} shows the
comparison between both approaches, discussing the magnetic field
distribution~\ref{ss:mag_struct}, then the deformation of the
star~\ref{ss:resu_deform} and studying the behavior when exploring
various compactnesses and EoSs~\ref{ss:resu_eos}. Finally,
Sec.~\ref{s:conc} summarizes the results and give some concluding
remarks. Some details about the integration of the perturbed solution
are given in the appendix. 

We use units such that $G=c=1$; units for electromagnetic quantities
shall be discussed in Sec.~\ref{ss:pert_mag}; metric signature is
$\left( -1, 1, 1, 1 \right)$.

\section{Perturbations of spherically symmetric
  configurations\label{s:pert}}
The overall idea here is to consider a spherically non-rotating
neutron star, perturbed by an axisymmetric 4-current distribution
$J_\mu = (0, 0, 0, J_\ph)$ , which induces a poloidal magnetic field
$B_\mu = (0, B_r, B_\tet, 0)$ deforming the star due to the
corresponding contribution to the energy-momentum tensor.  As long as
this deformation is supposed to be small, a perturbative approach in
the spirit of the so-called \textsl{slow rotation} approximation
devised by \cite{hartle-67, hartle-68} is applicable.  Perturbation
equations have been computed in Ref.~\cite{konno-99} and we recall the
main results here. All the ordinary differential equations presented
in this section are solved by a simple second-order Runge-Kutta
integrator, although special care is given to the treatment of the
origin at $r=0$, see App.~\ref{s:app_dl} and of the surface of the
star at $r=R$, see App.~\ref{s:app_match}.

\subsection{Background solution}\label{ss:background}
The background solution is that of a non-rotating and spherically
symmetric neutron star, whose metric $g_{\mu\nu}^{(0)}$ is
described by the well-known line element (remember that we use $c=1$)
\begin{equation}
  \label{e:metricTOV}
  ds^2 = g_{\mu\nu}^{(0)} dx^\mu dx^\nu
  = -e^\nu\, dt^2 + e^\la\, dr^2 + r^2\left( d\tet^2 + \sin^2\tet d\ph^2 \right),
\end{equation}
with $\nu(r)$ and $\la(r)$ functions of $r$ only. The star is assumed
to be composed of a perfect fluid, with the energy-momentum tensor
given by
\begin{equation}
  \label{e:perf_fluid}
  T^{\mu\nu} = (e + p)u^\mu u^\nu + p g^{\mu\nu},
\end{equation}
with $e$ the total fluid energy density, including rest-mass, $p$ its
pressure and $u^\mu$ its 4-velocity. Writing
\begin{equation}
  \label{e:def_m}
  e^{-\la(r)} = 1 - \frac{2m(r)}{r}
\end{equation}
and defining the log-enthalpy $h$ as
\begin{equation}
  \label{e:def_h}
  h(r) = \ln \left( \frac{e(r)+p(r)}{n_B(r)} \right),
\end{equation}
with the fluid baryon density $n_B(r)$, the Einstein field equation
can be written as the Tolman-Oppenheimer-Volkoff (TOV) system
\begin{eqnarray}
  \label{e:TOV}
  \frac{d m(r)}{dr} &=& 4\pi r^2 e(r), \\
  \frac{d h(r)}{dr} &=& -e^\la \left( \frac{m(r)}{r^2} + 4\pi r p(r)
                 \right). \nonumber 
\end{eqnarray}
This system is closed by an EoS prescribing
\begin{equation}
  \label{e:eos}
  p(h) \textrm{ and } e(h).
\end{equation}
The EoS models used in this study shall be discussed more in details
in Sec.~\ref{ss:resu_eos}. The metric potential $\nu(r)$ is determined by
\begin{equation}
  \label{e:dnudr}
    \frac{d\nu}{dr} = -\frac{2}{p(r)+e(r)}\frac{d p(r)}{dr}=-2\frac{d h(r)}{dr}.
\end{equation}
Selecting an EoS and choosing a value of central log-enthalpy $h(r=0) = h_c$, 
one can integrate the TOV system from the center to the surface $r=R$, where $p(R)=0$\footnote{In practice, for microphysically motivated EoS, the pressure does not vanish at the surface, but reaches a very small value, typically more than ten orders of magnitude below the central values. Thereby at the surface the crystal in the outer crust is composed of $^{56}$Fe corresponding to baryon number densities below a few times $10^{-9}$ fm$^{-3}$.},
and match $\la(r)$ and $\nu(r)$, see App.~\ref{s:app_match} for details, to the
Schwarzschild vacuum solution
\begin{eqnarray}
  \label{e:TOV_ext}
  m_{\rm vac}(r) &=& M,\\
  \nu_{\rm vac}(r) &=& \ln \left( 1 - \frac{2M}{r} \right). \nonumber
\end{eqnarray}
This determines the background star's mass $M$ and radius $R$.
With the matching condition~(\ref{e:nuDetla}) Eq.~(\ref{e:dnudr}) can be integrated
\begin{equation}
  \label{e:nuhsol}
    \nu(r) =\nu_{\rm vac}(R) - 2h(r),
\end{equation}
connecting the $g_{tt}$ metric potential $\nu(r)$ for $r\le R$ 
directly to the log-enthalpy $h(r)$.

Both from an analytical and numerical perspective the system of equations~(\ref{e:TOV}) has some practical shortcomings: the coordinate
singularity at $r=0$ has to be treated with care and the integration 
domain, i.e. the radius of the star $r=R$ is not known prior to numerically
solving Eqs.~(\ref{e:TOV}). An alternative formulation using the 
log-enthalpy $h$ as integration variable put forward in Ref.~\cite{lindblom-92}:
\begin{eqnarray}
  \label{e:TOVr2z}
  \frac{d r^2(h)}{d h} &=& - \frac{2 r(h)^2(1-2x(h))}{4\pi r(h)^2 p(h)+x(h)},\\
  \frac{d x(h)}{d h} &=& \left(2\pi e(h) -\frac{x(h)}{2r(h)^2}\right)\frac{d r^2(h)}{d h},\nonumber
\end{eqnarray}
with $x(h)=m(h)/r(h)$, is better suited for analytical expansions at 
the star's center and is numerically both easier to implement and more stable, when
compared to Eqs.~(\ref{e:TOV}).

\subsection{Magnetic field}\label{ss:pert_mag}
As stated above, we assume an axisymmetric current distribution
$J_\mu(r,\tet)$ and no rotation, which implies no electric field and that the
electromagnetic 4-potential can be written as $A_\mu = \left(0, 0, 0,
A_\ph(r, \tet)\right)$ (see e.g.~Refs.\cite{bonazzola-93,bocquet-95,konno-99}). 
The Maxwell equation then takes the form
\begin{equation}
  \label{e:Maxwell}
  e^{-\la}\dder{A_\ph}{r} + \frac{1}{2}\left( \nu' - \la'\right)
  e^{-\la}\der{A_\ph}{r} + \frac{1}{r^2} \dder{A_\ph}{\tet} -
  \frac{1}{r^2}\cot \tet \der{A_\ph}{\tet} = 4\pi J_\ph.
\end{equation}
Note here that we choose units such that the vacuum magnetic
permeability $\mu_0 = 4\pi$.

Following Ref.~\cite{konno-99}, we expand $A_i$ and $J_i$ onto vector
spherical harmonics and \textsl{consider only a dipole magnetic field},
$\ell=1$. Then, the differential equation to be solved for $a_1$, the
$\ell=1$ component of the electromagnetic 4-potential, is
\begin{equation}
  \label{e:Maxwell_l1}
  \frac{d^2a_1}{dr^2} + \frac{1}{2}\left( \nu'- \la' \right)
  \frac{da_1}{dr} - 2 e^\la \frac{a_1}{r^2} = 4\pi e^\la j_1.
\end{equation}
A star in hydrostatic equilibrium must adhere to the relativistic Euler 
equation including in this case Lorentz force terms, see e.g.
Refs.~\cite{colaiuda-08,ioka-03,ioka-04} for details. In this context an 
integrability condition~\cite{konno-99} can be derived putting energy density, pressure, 
and the $\ell=1$ component of the current distribution into relation:
\begin{equation}
  \label{e:j1_mat}
  j_1(r) = c_0 r^2 \left( e(r) + p(r) \right),
\end{equation}
with $c_0$ an arbitrary constant, determining the current amplitude
and thus, the magnetic field strength. Note that this form of the
current distribution is compatible with the more general form of
stationary, axisymmetric and circular ones discussed in
Ref.~\cite{bonazzola-93}, in the specific case without rotation. In particular, the assumption of circularity prevents the existence of
meridional currents and should be relaxed if one wishes to study a
more realistic magnetar setting with toroidal currents and magnetic
field. 

Equation~(\ref{e:Maxwell_l1}) is integrated from the center to the
surface of the star, where the function $a_1(r)$ is matched to the
vacuum solution:
\begin{equation}
  \label{e:a1_vacuum}
  a_1^{\rm vac}(r) = -\frac{3\mu}{8M^3}r^2 \left[ \ln \left(1 -
      \frac{2M}{r} \right) + \frac{2M}{r} + \frac{2M^2}{r^2} \right].
\end{equation}
Asymptotically, the constant $\mu$ can be identified as the
star's magnetic dipole moment. In practice, Eq.~(\ref{e:Maxwell_l1})
is solved with $a_1(0) = a_1'(0) = 0$ twice: first to get a particular
solution and then, setting its right-hand side to zero to get a
homogeneous one. The coefficient multiplying the homogeneous solution,
and the magnetic moment $\mu$ are determined by enforcing continuity
of both $a_1(r)$ and $a_1'(r)$ at $r=R$.

The magnetic field is then deduced in the whole star as
\begin{equation}
  \label{e:def_B}
  B_{\hat{\mu}} = \left( 0, -\frac{2 a_1}{r^2}\cos \tet, \frac{e^{-\la/2}
      a_1'}{r}\sin \tet, 0 \right),
\end{equation}
where the components are expressed in the normalized vector basis
(tetrad). Particular values are the polar, equatorial and central
magnetic field values, given respectively by:
\begin{eqnarray}
  B_{\rm pole} &=& -\frac{2 a_1(R)}{R^2} \nonumber,\\
  B_{\rm eq} &=& \frac{e^{-\la(R)}a_1'(R)}{R},   \label{e:B_part}\\
  B_c &=& -2\alpha_0 \nonumber,
\end{eqnarray}
with $\alpha_0 = \lim_{r\to0} a_1(r)/r^2$ (see also App.~\ref{s:app_dl}).
By construction the perturbative approach presented in this subsection is 
linear: the magnetic field $B_{\hat{\mu}}$, the potential $a_1$, and 
the current $j_1$ are all linear in $c_0$ or respectively $\alpha_0$, which 
by means of Eqs.~(\ref{e:B_part}) can be translated into a linear dependence on 
$B_c$.

\subsection{Deformation induced by the magnetic
  field}\label{ss:deform}
Due to the magnetic field $\ell=1$ component described above,
spacetime is perturbed and is no longer spherically symmetric. The
metric is given by (see Ref.~\cite{konno-99} and Ref.~\cite{hartle-67} for
details)
\begin{eqnarray}
  ds^2 &=& -e^{\nu(r)} \left[ 1 + 2\left( h_0(r) + h_2(r)P_2(\cos\tet )
    \right) \right] dt^2 \nonumber \\
  && + e^{\la(r)} \left[ 1 + \frac{2 e^{\la(r)}}{r} \left( m_0(r) +
      m_2(r)P_2(\cos \tet) \right) \right] dr^2 \nonumber \\
  && + r^2 \left[ 1 + 2 k_2(r) P_2(\cos \tet) \right] \left( d\tet^2 +
r^2 \sin^2 \tet\, d\ph^2 \right),   \label{e:pert_metric}
\end{eqnarray}
where $P_2(x)$ is the Legendre polynomial of degree~2, and
$h_0, h_2, m_0, m_2$ and $k_2$ are functions of $r$ only, describing
the corrections to the second order in magnetic field amplitude. They
can be computed by solving the perturbed Einstein field equation,
taking into account the contribution from the magnetic field in the
energy-momentum tensor.

As we are interested in the deformation of the star, we focus on the
functions $h_2(r)$ and $k_2(r)$. Following Ref.~\cite{colaiuda-08}, defining
\begin{equation}
  y_2(r) = h_2 + k_2 - \frac{e^{-\la}}{6} a_1'^2 -
  \frac{2e^{-\la}}{3r}a_1 a_1' - \frac{2}{3r^2}a_1^2, \label{e:def_y2}
\end{equation}
one obtains the following set of equations for $h_2(r)$ and $y_2(r)$
\begin{eqnarray}
  h_2' &=& -\frac{4e^\la}{\nu' r^2}\, y_2
  + \left[ \frac{8\pi e^\la}{\nu'} \left( e + p \right) -
      \frac{2}{r^2\nu'} \left( e^\la - 1 \right) -\nu' \right]\,
      h_2 \nonumber \\
  &&+ \frac{\nu'}{3} e^{-\la} a_1'^2 + \frac{4}{3r^2}a_1 a_1' -
      \frac{16\pi}{3\nu'r^2} e^\la j_1 a_1, \label{e:h2} \\
  y_2' &=& -\nu'\, h_2 + \frac{\nu'}{2} e^{-\la} a_1'^2 + \left[
                         \frac{e^{-\la}}{r} \left( \nu' + \la' +
                         \frac{2}{r} \right) - \frac{2}{r^2} \right]
                         \frac{a_1a_1'}{3} \nonumber \\
   && - \frac{4\pi}{3}j_1 \left( a_1' + \frac{2a_1}{r} \right). \label{e:y2}
\end{eqnarray}
Similarly to the procedure for the magnetic potential described in Sec.~\ref{ss:pert_mag}, 
the system~(\ref{e:h2})--(\ref{e:y2}) is integrated twice on the interval $0 \leq r \leq
R$ (behaviors for $h_2$ and $y_2$ around $r=0$ are described in
App.~\ref{s:app_dl}): first, a couple of particular solutions is
obtained with the full system, then a couple of homogeneous ones is
computed setting $a_1=j_1=0$. Finally, these interior solutions are
matched to exterior, vacuum solutions~\cite{konno-99, colaiuda-08}:
\begin{eqnarray}
  h_2^{\rm vac} &=& K\, Q_2^2(z) + \hat{h}_2(z), \label{e:h2_vac} \\
  y_2^{\rm vac} &=& - \frac{2K}{\sqrt{z^2-1}} Q_2^1(z) + \hat{y}_2(z)
                    - \frac{e^{-\la}}{6} (a_1')^2
                    -\frac{2}{3r}e^{-\la}a_1 a_1' -
                    \frac{2(a_1)^2}{3r^2} \label{e:y2_vac},
\end{eqnarray}
with $K$ being a constant to be determined by this matching.
\begin{eqnarray}
  Q_2^2(z) &=& \frac{z \left( 5 - 3z^2 \right)}{z^2 - 1} + \frac{3}{2}
  \left( z^2 - 1 \right) \ln \left( \frac{z+1}{z-1} \right),
  \nonumber\\
  Q_2^1(z) &=& \frac{2 - 3z^2}{\sqrt{z^2-1}} + \frac{3}{2}z\left(
               \sqrt{z^2 - 1} \right) \ln \left( \frac{z+1}{z-1}
               \right) \label{e:assoc_leg}
\end{eqnarray}
are associated Legendre functions of the second kind,
\begin{equation}
  \label{e:def_z}
  z = \frac{r}{M} - 1,
\end{equation}
and
\begin{eqnarray}
  \hat{h}_2 & =& -\frac{3\mu^2}{16M^4} \left( 3z - \frac{4z^2 +
                 2z}{z^2 -1} \right) - \frac{3\mu^2}{32M^4} \left(
                 3z^2 - 8z -3 - \frac{8}{z^2 - 1} \right) \ln
                 \left(\frac{z-1}{z+1} \right) \nonumber\\
            && + \frac{3\mu^2}{16M^4}\left( z^2 -1 \right) \ln\left(
               \frac{z-1}{z+1} \right)^2, \label{e:hat_h2}\\
  \hat{y}_2 &=& \frac{3\mu^2}{8M^4} \frac{7z^2-4}{z^2-1} +
                \frac{3\mu^2}{16M^4} \frac{z \left( 11z^2 -7
                \right)}{z^2 -1} \ln \left( \frac{z-1}{z+1} \right)
                \nonumber \\
            &&+ \frac{3\mu^2}{16M^4} \left( 2z^2 + 1 \right) 
                \ln \left(\frac{z-1}{z+1} \right)^2 \label{e:hat_y2}.
\end{eqnarray}

Being solutions of the perturbed Einstein equations, both metric
perturbations $h_2$ and $y_2$ take into account contribution of the
magnetic field to the energy-momentum tensor. This can be seen in the
system (\ref{e:h2})--(\ref{e:y2}), where quadratic terms in $a_1$ and
its derivative appear. Outside the star, these are nonzero and
represent the magnetic field influence on the perturbed metric in
vacuum.

Once $h_2(r)$ and $y_2(r)$ are computed, it is possible to deduce the
ellipticity of the star~\cite{chandrasekhar-74, konno-99}:
\begin{equation}
  \label{e:def_ellip}
  \varepsilon_{\rm surf} = \frac{3h_2(R)}{R\nu'(R)} - \frac{2c_0
    a_1(R)}{R\nu'(R)} - \frac{3k_2(R)}{2}.
\end{equation}
The first term in this equation can be interpreted as the deformation
from the interaction of the currents with the magnetic field (``Lorentz
force''); the second one comes from the perturbation of the
gravitational field from magnetic stresses and the last one was
described as a ``purely relativistic term'' arising from the
definition of the circumferential radius~\cite{konno-99}. 
Then, one can define the star's mass quadrupole moment
\begin{equation}
  \label{e:def_Q}
  Q =  \frac{8M^3}{5}K - \frac{6\mu^2}{5M},
\end{equation}
which can be used to define the star's quadrupole
ellipticity
\begin{equation}
  \label{e:def_eQ}  
  \varepsilon_Q = - \frac{Q}{I},
\end{equation}
where $I$ is the moment of inertia of the background (non-magnetized)
star, and can be computed within slow-rotation
approximation~\cite{hartle-68}. The respective properties of
$\varepsilon_{\rm surf}$ and $\varepsilon_Q$ have been discussed in
Ref.~\cite{colaiuda-08}. Note again that by construction the functions $h_2(r)$,
$k_2(r)$, $y_2(r)$, and hence $\varepsilon_{\rm surf}$ and $Q$ are
quadratic in the magnetic field.

\section{Full numerical solutions\label{s:magstar}}
In this section we present the framework and assumptions made to compute
numerical models of magnetized neutron stars, including the
possibility of strong deformation, using the code
\texttt{magstar}~\cite{bocquet-95, novak-03} and the \textsc{lorene}
library~\cite{lorene}. 

\subsection{Stationary axisymmetric spacetime and global quantities}\label{ss:axi}
Following the approach put forward in Ref.~\cite{bonazzola-93}, we
assume spacetime to be stationary, axisymmetric and the
energy-momentum tensor to fulfill the circularity condition. This
implies that the magnetic field configuration to be either purely
toroidal or purely poloidal, and we assume the latter. Moreover, if we
assume no electric field, the 4-current compatible with these
hypotheses has the following form $J^\mu = \left( 0, 0, 0, J^\ph
\right)$. The electromagnetic 4-potential $A_\mu$ takes exactly the
same form as in Sec.~\ref{ss:pert_mag}: $A_\mu = \left(0, 0, 0,
A_\ph(r, \tet)\right)$. On the other hand, the metric line element is
different from that of Eq.~(\ref{e:metricTOV}) because of a different
choice of coordinates, and from that of Ref.~\cite{bonazzola-93}
because we consider only non-rotating stars, without any electric
field\footnote{there is thus no frame-dragging ``shift'' term.}
\begin{equation}
  \label{e:metricMSQI}
  ds^2 = -N^2 dt^2 + A^2 \left( dr^2 + r^2 d\tet^2 \right) + B^2r^2
  \sin^2 \tet d\ph^2 ,
\end{equation}
where all three metric potentials $N, A, B$ are functions of $(r,
\tet)$ only. Note here that the coordinates in Eq.~(\ref{e:metricMSQI})
are different from those of Eq.~(\ref{e:metricTOV}), but we will make
use of the same notations. This point shall be discussed again when
comparing both approaches in Sec.~\ref{s:compare}.

The electromagnetic field tensor is defined, as usual as $F_{\mu\nu} =
\partial_\mu A_\nu - \partial_\nu A_\mu$ and thus the magnetic field
measured by the Eulerian observer is given by
\begin{equation}
  \label{e:B_num}
  B_{\hat{\mu}} = \left( 0, \frac{1}{AB r^2 \sin \tet}
    \der{A_\ph}{\tet}, -\frac{1}{AB r \sin \tet}\der{A_\ph}{r}, 0 \right).
\end{equation}
The magnetic dipole moment of the star $\mu$ is determined as the leading
term of the asymptotic behavior of this magnetic field
\begin{equation}
  \label{e:def_mu_num}
  B_{\hat{r}} \underset{r\to\infty}{\sim} \frac{2\mu\cos \tet}{r^3}
  \quad \textrm{ and } \quad B_{\hat{\tet}} \underset{r\to\infty}{\sim}
  \frac{\mu\sin\tet}{r^3}. 
\end{equation}
Similarly, the total mass of the star $M$ and the mass quadrupole
moment $Q$ can be obtained from the asymptotic behavior at spatial
infinity of some metric coefficients, i.e. the leading terms
\begin{eqnarray}
  \ln N &\underset{r\to\infty}{=}& -\frac{M}{r} +
\frac{b}{3}\left(\frac{M}{r} \right)^3 - \frac{\bar{Q}}{r^3} P_2(\cos
\tet ) + O\left( \frac{1}{r^4} \right) , \label{e:asym_N}\\
  N\,B  &\underset{r\to\infty}{=}& 1 + b\left( \frac{M}{r} \right)^2 +
                                   O\left( \frac{1}{r^4} \right). \label{e:asym_NB}
\end{eqnarray}
$\bar{Q}$ and $b$ are combined to compute the mass quadrupole moment
$Q$ following the prescription by Ref.~\cite{pappas-12} (see
also Ref.~\cite{bonazzola-96,friedman-13})
\begin{equation}
  \label{e:Q_magstar}
  Q = \bar{Q} - \frac{4}{3}\left( b + \frac{1}{4}\right) M^3.
\end{equation}
Note that the computation of these asymptotic coefficients can
eventually be done through volume integrals (see~e.g. Refs.~\cite{bonazzola-93,
  prix-05} for details).

Finally, gauge-independent radii $R_{\rm circ}(\tet)$ are defined as a
function of the co-latitude $\tet$ as the measured length on the
star's surface of a circle around the $z$-axis at this co-latitude,
divided by $2\pi$. This gives
\begin{equation}
  \label{e:rcirc}
  R_{\rm circ}\left( \tet \right) = B^2\left(r_s(\tet), \tet \right) r_s(\tet),
\end{equation}
where $r_s(\tet)$ is the coordinate radius of the surface of the
star. For $\tet=\pi/2$ the usual circumferential equatorial radius is
recovered. Note that for $\tet \to 0$ the limit of the expression
(\ref{e:rcirc}) is well-defined. Therefore, the ellipticity of the
star, cf. Eq.~(\ref{e:def_ellip}), shall be computed as
\begin{equation}
  \label{e:ellip_num}
  \varepsilon_{\rm surf} = \frac{R_{\rm circ}\left( \pi/2 \right) -
    R_{\rm circ}(0)}{R_{\rm circ}(\pi/2)}.
\end{equation}

\subsection{Einstein-Maxwell and equilibrium equations}\label{ss:num_eqs}

The energy-momentum tensor takes the form of that of a perfect
fluid~(\ref{e:perf_fluid}), with the additional contribution of the
electromagnetic field $T_{\mu\nu}^{\rm EM} = 1/(4\pi) \left(
  F^{\mu\sigma} F^{\nu}_{\sigma} - 1/4 F_{\sigma \rho} F^{\sigma \rho}
  g^{\mu\nu} \right)$.

A first equation to be solved is that of the conservation of
energy-momentum, which can be written in the case of a non-rotating
star as
\begin{equation}
  \label{e:div_Tmunu}
  (e + p) \left( \der{h}{x^i} + \der{\ln N}{x^i} \right) - F_{i\sigma}
  J^\sigma = 0,
\end{equation}
where the last term can be interpreted as the Lorentz force and is
written as $F_{i\ph}J^\ph = \der{A_\ph}{x^i}\, J^\ph$. The
integrability condition is thus (see~Ref.~\cite{bonazzola-93})
\begin{equation}
  \label{e:current_func}
  F_{i\sigma} J^\sigma = -\left(e + p\right) \der{\Phi}{x^i},
\end{equation}
with
\begin{equation}
  \label{e:def_Phi}
  \Phi(r, \tet) = - \int_0^{A_\ph(r,\tet)} f(x) dx,
\end{equation}
where $f$ is the \textsl{current function}, such that
\begin{equation}
  \label{e:def_current_func}
  J^\ph = \left( e+p \right) f\left( A_\ph \right). 
\end{equation}
Note that $f$ can be chosen arbitrarily and, to be consistent with
the condition~(\ref{e:j1_mat}) used in the perturbative approach, we
simply write $f(x) = c_0$. Thus, the first integral takes the simple
form
\begin{equation}
  \label{e:first_integral}
  \ln h(r,\tet) + \ln N(r, \tet) + \Phi(r, \tet) = \textrm{const}.
\end{equation}

The Maxwell equations reduce in the present setup to only one
non-vanishing relation, i.e. the Maxwell-Amp\`ere equation
\begin{equation}
  \tilde{\Delta}_3 \left( \frac{A_\ph}{r\sin\tet} \right)
  = 4\pi A^2 B^2 J^\ph r \sin \tet  + \frac{N}{B r \sin \tet}
  \partial A_\ph \partial \left( \frac{B}{N} \right), \label{e:maxwell}
\end{equation}
and the Einstein equations manifest in a set of three elliptic partial
differential equations for the metric potentials defined in
Eq.~(\ref{e:metricMSQI})
\begin{eqnarray}
  \Delta_3 N &=& 4\pi N A^2 \left( \mathcal{E} + \mathcal{S}^i_{\,i}
                 \right) - \frac{1}{B} \partial N
                 \partial B, \label{e:deltaN} \\
  \Delta_2 \left[ \left( NB-1 \right) r\sin\tet \right]
             &=& 16\pi N A^2 B r \sin \tet \left( \mathcal{S}^r_{\, r} +
                 \mathcal{S}^\tet_{\, \tet} \right), \label{e:deltaNB}\\
  \Delta_2 \left( NA \right)
             &=& 8\pi N A^3 \mathcal{S}^\ph_{\, \ph} +
                 \frac{N}{A} \left( \partial A \right)^2, \label{e:deltaNA}
\end{eqnarray}
with the following notations:
\begin{eqnarray}
  \Delta_2 &=& \dder{}{r} + \frac{1}{r}\der{}{r} + \frac{1}{r^2}
               \dder{}{\tet}, \nonumber\\
  \Delta_3 &=& \dder{}{r} + \frac{2}{r}\der{}{r} + \frac{1}{r^2}
               \dder{}{\tet} + \frac{1}{r \tan \tet} \der{}{\tet},
               \nonumber \\
  \tilde{\Delta}_3 &=& \dder{}{r} + \frac{2}{r}\der{}{r} + \frac{1}{r^2}
                       \dder{}{\tet} + \frac{1}{r \tan \tet} \der{}{\tet}
                       - \frac{1}{r^2 \sin^2 \tet},
               \nonumber \\
  \partial X \partial Y &=& \der{X}{r}\der{Y}{r} + \frac{1}{r^2}
                            \der{X}{\tet} \der{Y}{\tet} \nonumber .
\end{eqnarray}
The quantities $\mathcal{E}$ and $\mathcal{S}^i_{\, j}$ are
obtained from the 3+1 decomposition of the energy-momentum tensor (see
e.g. Ref.~\cite{gourgoulhon-12}). In our case of a perfect fluid with a
magnetic field, one has~\cite{bonazzola-93}:
\begin{eqnarray}
  \mathcal{E} &=& e + \frac{1}{8\pi}  \left( B^r B_r + B^\tet B_\tet
                            \right), \label{e:e_euler} \\
  \mathcal{S}^r_{\, r} &=& p + \frac{1}{8\pi} \left( B^\tet B_\tet -
                           B^r B_r \right), \label{e:s_euler_r} \\
  \mathcal{S}^\tet_{\, \tet} &=& p + \frac{1}{8\pi} \left( B^r B_r - B^\tet B_\tet 
                                 \right), \label{e:s_euler_th} \\
  \mathcal{S}^\ph_{\, \ph} &=& p + \frac{1}{8\pi} \left( B^r B_r + B^\tet B_\tet
                            \right). \label{e:s_euler} \label{e:s_euler_ph}
\end{eqnarray}
Note here that, because of the presence of the magnetic field, the
constraint tensor $\mathcal{S}^i_{\, j}$ is not isotropic and the three
diagonal components are different one from another. It is thus not
possible to define any type of scalar magnetic pressure, not even with
``parallel'' and ``perpendicular'' components as very often proposed in the literature.  

\subsection{Numerical strategy and checks}\label{ss:num_tests}

A numerical model is obtained by fixing the EoS, the value of the
central log-enthalpy $h_c$ and the constant defining the current
function~(\ref{e:def_current_func}) $c_0$. Poisson-like equations
(\ref{e:maxwell})--(\ref{e:deltaNA}) are solved by so-called \textsl{spectral
  methods}~\cite{grandclement-09, grandclement-01}, where all fields are
represented by truncated series of Chebyshev polynomials for the
coordinate $r$, and Fourier series in $\tet$. Because the sources of
these equations extend up to spatial infinity, we make use of a
multi-domain decomposition for the coordinate $r$, with a first domain
dealing with the coordinate singularity at $r=0$, then a certain
number of shells, and finally a compactified domain where the
coordinate is changed to $u = 1/r$ and the spectral decomposition is
made in terms of $u$ mapped to the interval $[-1, 1]$. This system is
completed by the first integral giving the
equilibrium~(\ref{e:first_integral}) and the EoS.

The system to be solved being nonlinear, a \textsl{fixed-point
  iteration} is used, with a first guess being a flat metric and a
parabolic log-enthalpy density. At each iteration step, the source
terms of the Poisson equations~(\ref{e:maxwell})--(\ref{e:deltaNA})
are fixed and the linear part of the system is solved with the
inversion of the Laplace operators, which gives new values for
$N, A, B$ and $A_\ph$. $h$ is obtained from the equilibrium
condition~(\ref{e:first_integral}), and the EoS gives $e$ and $p$ to
compute updated source terms. This is done for several iterations,
with some relaxation, until the relative difference between two
successive log-enthalpy profiles becomes smaller than a given
threshold ($10^{-10}$ in this study). Finally, the accuracy of the
solution is checked using the so-called \textsl{virial
  identities}~\cite{gourgoulhon-94, bonazzola-94}, which are global
(integral) relations including metric and energy-momentum, that any
stationary spacetime should fulfill. These tests, called GRV2 and GRV3
are independent from the equations that are solved and give an upper
dimensionless bound on the relative errors on the numerical solution. With
microphysical EoSs (see Sec.~\ref{ss:resu_eos}), we have checked that
both these error indicators always remained below $10^{-4}$.

\section{Comparison of both approaches}\label{s:compare}
In this section, we will study two features of magnetized neutron
stars: the magnetic field distribution and the deformation it
induces. Within our model, the magnetic field is assumed to be purely
poloidal but the perturbative approach considers only a dipolar
distribution, whereas the fully numerical one allows for higher
multipoles. However, when comparing perturbative (following the work by Konno et
al.~\cite{konno-99}) and fully numerical (\texttt{magstar} code)
approaches, one must be careful in considering gauge-independent
quantities and, in particular the $(r,\tet)$ coordinates defined by
the line element~(\ref{e:pert_metric}) are different from those used
by the numerical code~(\ref{e:metricMSQI}). Nevertheless, physically
defined points can be used, such as the star's center, the poles or the
equator. In what follows we compare quantities at given polar magnetic
field values $B_{\rm pole} = B_{\hat{r}}$ at the star's surface and
$\tet=0$. In Secs.~\ref{ss:mag_struct} and \ref{ss:resu_deform},
results are given using the DDFGOS(APR) EoS~\cite{davis-25}, which is
shortly noted APR and further discussed in Sec.~\ref{ss:resu_eos}. 

\subsection{Magnetic field structure}\label{ss:mag_struct}
\begin{figure}[ht]
  \centering
  \subfloat[\centering]{\includegraphics[width=8.1cm]{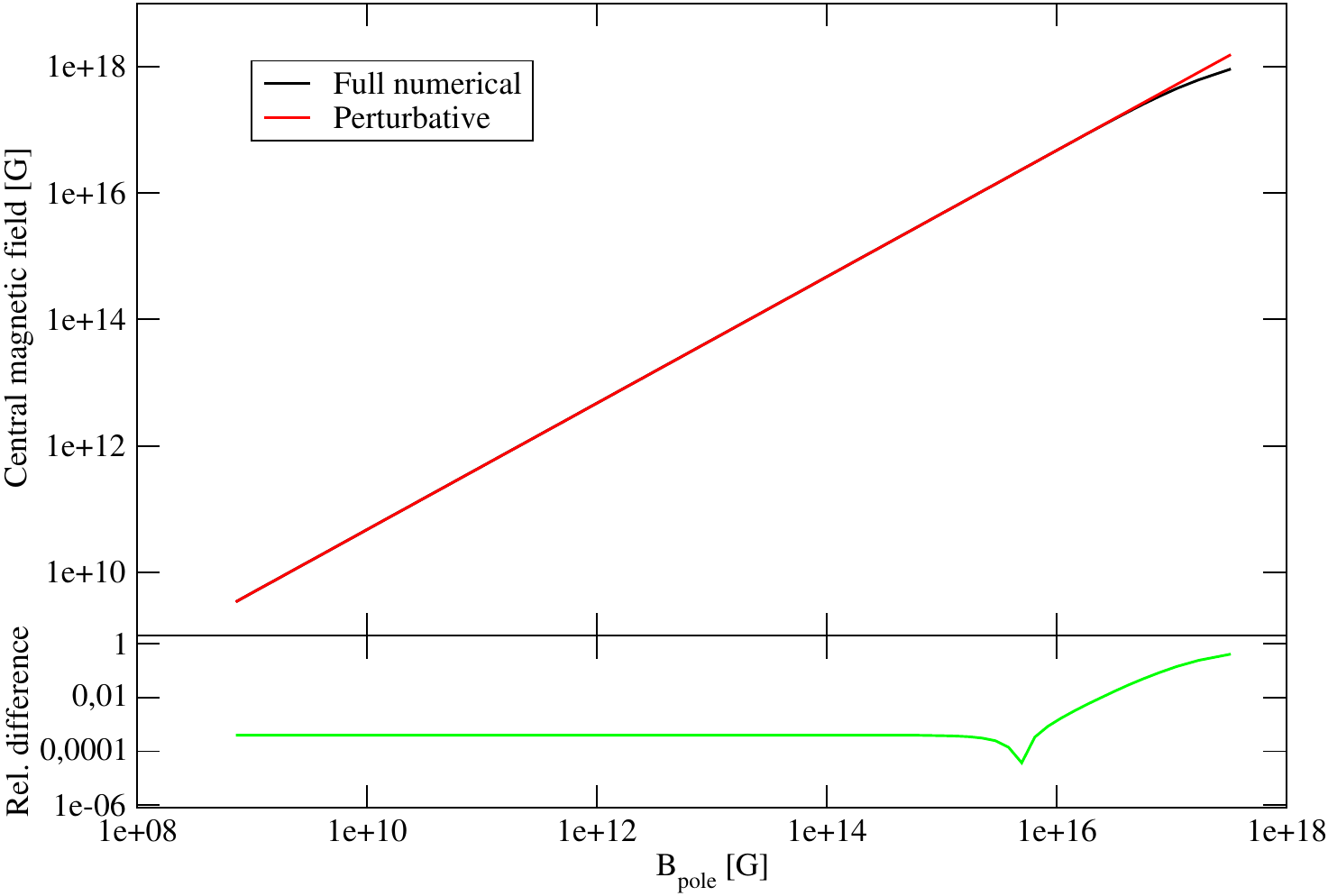}}
  \subfloat[\centering]{\includegraphics[width=8.1cm]{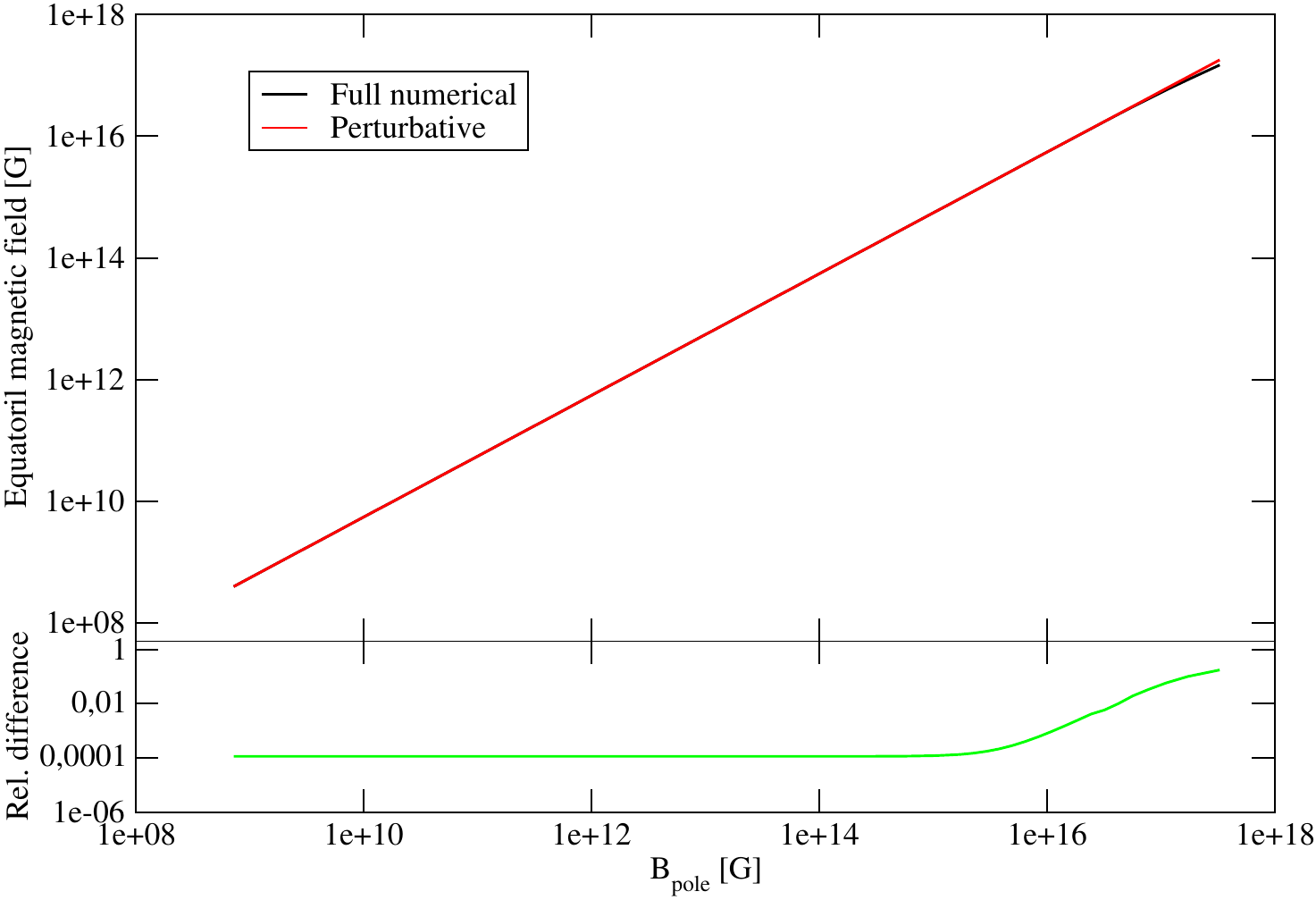}}
  \caption{Central and equatorial magnetic fields as functions of the
    polar one, computed with the perturbative approach or in a full
    numerical one. In the bottom are shown the relative differences
    between both methods. The background non-magnetized star has a
    compactness of $M/R=0.15$ and the DDFGOS(APR) EoS with unified crust is used.}
  \label{f:magfields}
\end{figure}

In Figs.~\ref{f:magfields} the magnetic field values  are plotted at
the star's center and at the equator for increasing $B_{\rm pole}$,
either computed with the perturbative approach of
Sec.~\ref{ss:pert_mag} or using the \texttt{magstar} code described in 
Sec.~\ref{s:magstar}. As expected, for relatively low magnetic fields
($B_{\rm pole} \lesssim 10^{16}$~G), the difference is very small, and it 
starts to deviate more significantly close to the highest value of $B_{\rm pole}$ the
star can support, where the difference is about 100\%. This behavior
indicates that both assumptions used in the perturbative approach,
namely the use of only dipolar currents and a spherical background star, are
suited for the description of a poloidal magnetic field. Of course,
as discussed in Sec.~\ref{ss:resu_eos}, the value of a
``threshold'' magnetic field, below which the perturbative approach is
valid, depends on both the EoS and the star's compactness.

\subsection{Deformation}\label{ss:resu_deform}
We now turn to the comparison of deformation indicators, with the mass
quadrupole moment $Q$ obtained either perturbatively from
Eq.~(\ref{e:def_Q}), or numerically from Eq.~(\ref{e:Q_magstar}), and
the surface ellipticity $\varepsilon_{\rm surf}$ using Eqs.~(\ref{e:def_ellip})
and (\ref{e:ellip_num}) for perturbative and numerical results respectively.
It would be completely equivalent to study the quadrupole ellipticity
$\varepsilon_Q$~(\ref{e:def_eQ}), since the moments of inertia $I$
computed perturbatively and numerically coincide up to six digits. 
\begin{figure}[ht]
  \centering
  \subfloat[\centering]{\includegraphics[width=8.1cm]{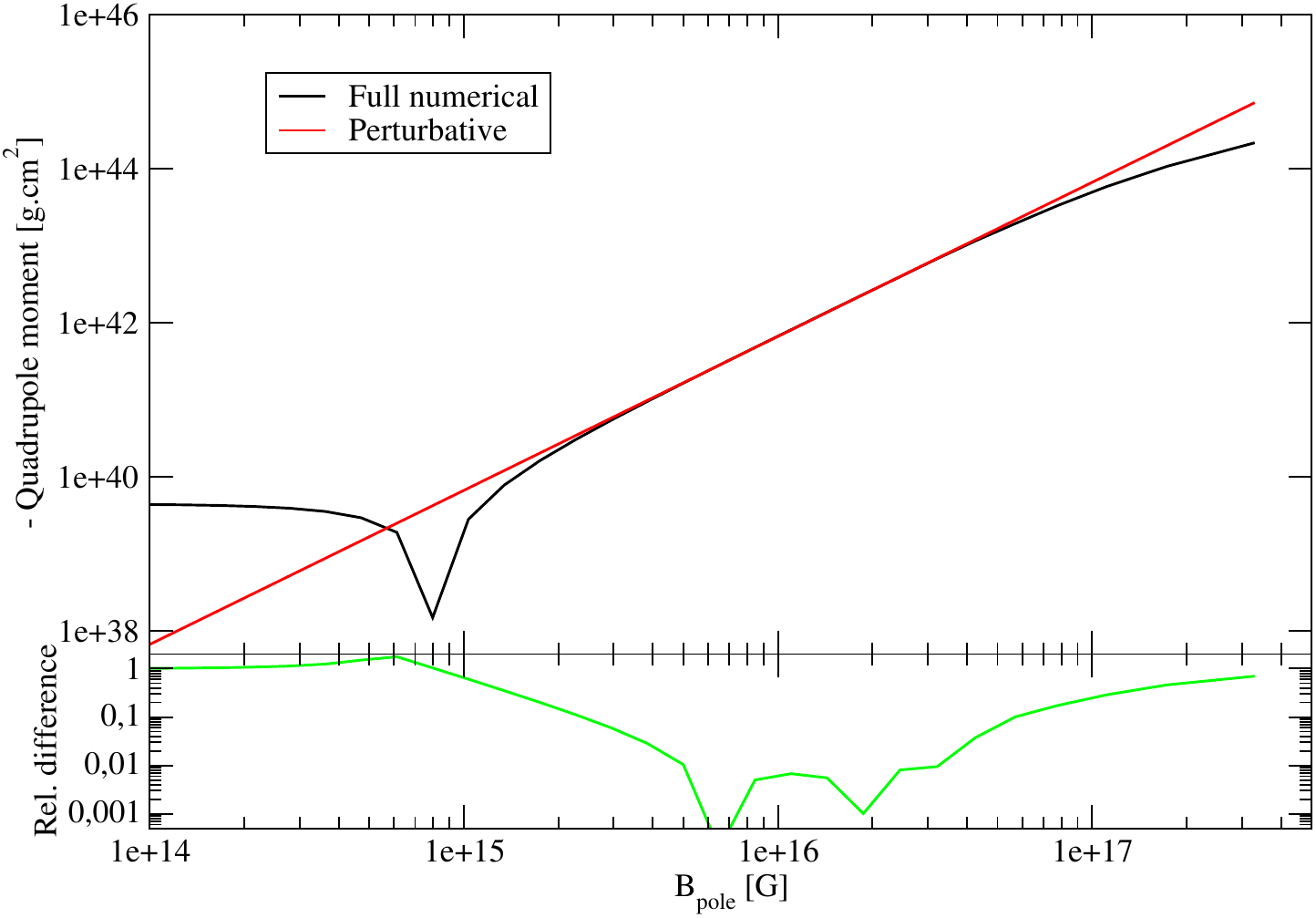}}
  \subfloat[\centering]{\includegraphics[width=8.1cm]{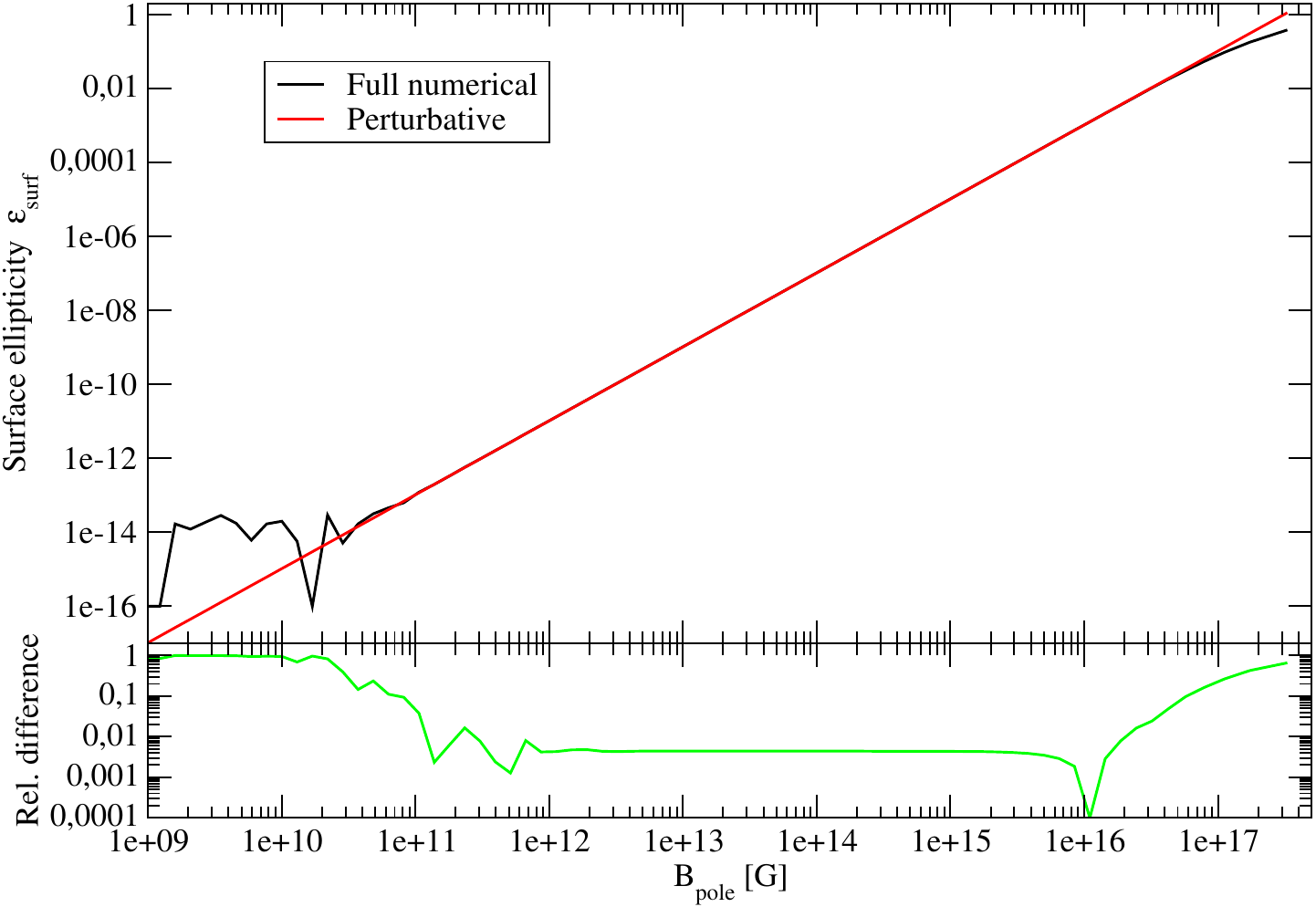}}
  \caption{Mass quadrupole momentum $Q$ of the star (left panel, with a minus
    sign) and surface ellipticity $\varepsilon_{\rm surf}$ (right panel) as functions of
    the polar magnetic field, computed with the perturbative approach
    or in a full numerical one. In the bottom are shown the relative
    differences between both methods. The background non-magnetized star has a
    compactness of $M/R=0.15$ and the DDFGOS(APR) EoS with unified crust is used.}
  \label{f:deform}
\end{figure}
$\varepsilon_{\rm surf}$ and $-Q$ are plotted in Figs.~\ref{f:deform},
computed both within the perturbative and numerical models,
with increasing polar magnetic field value. The difference with
the magnetic field distribution discussion in Sec.~\ref{ss:mag_struct} is
that both approaches diverge not only for high values of $B_{\rm
  pole}$, but at low ones, too. This can be understood because these
quantities are computed in the numerical code as differences between
numbers of similar values, as one can see from Eq.~(\ref{e:ellip_num})
for $\varepsilon_{\rm surf}$. In Eq.~(\ref{e:Q_magstar}), the mass
quadrupole moment is determined from asymptotic
expansions~(\ref{e:asym_N})-(\ref{e:asym_NB}), where the relevant terms are
sub-dominant. As a result, for relatively low polar magnetic field
values, the perturbative approach is more accurate than the full
numerical one. Nevertheless, there is room for improvement of the
numerical code in order to better compute these quantities.

As far as the high values of $B_{\rm pole}$ are concerned, the linear
approximation of the perturbative model breaks down as nonlinear
general-relativistic effects enter into place. We thus define two 
threshold values for the polar magnetic field $B_{\rm pole}$: a first
one ($B_{\rm pole}^5$) above which the relative difference in ellipticity
$\varepsilon_{\rm surf}$ is greater than 5\%, and the second one
($B_{\rm pole}^{50}$) above which it is greater than 50\%. For the
case shown in Fig.~\ref{f:deform}, one finds $B_{\rm pole}^5 =
4.3\times 10^{16}$~G and $B_{\rm pole}^{50} = 2.3\times
10^{17}$~G. This gives some idea about which values of the polar
magnetic field can be accessible with the perturbative approach and
which ones should be studied with the numerical code, depending on the
required accuracy.

\subsection{Dependence on the equation of state}\label{ss:resu_eos}
The values of $B_{\rm pole}^5$ and $B_{\rm pole}^{50}$ discussed above
depend in general on the particular background (non-magnetized) model
used in the perturbative approach. This last depends on two inputs:
the EoS and the central enthalpy $h_c$. We prefer to replace this last
quantity by the non-magnetized star's compactness $\mathcal{C} = M/R$,
whereas we have until now only considered a star with
$\mathcal{C}=0.15$ and one EoS model. In the following we vary both the compactness as well as the EoS.
\begin{figure}[ht]
  \centering
  \includegraphics[width=12cm]{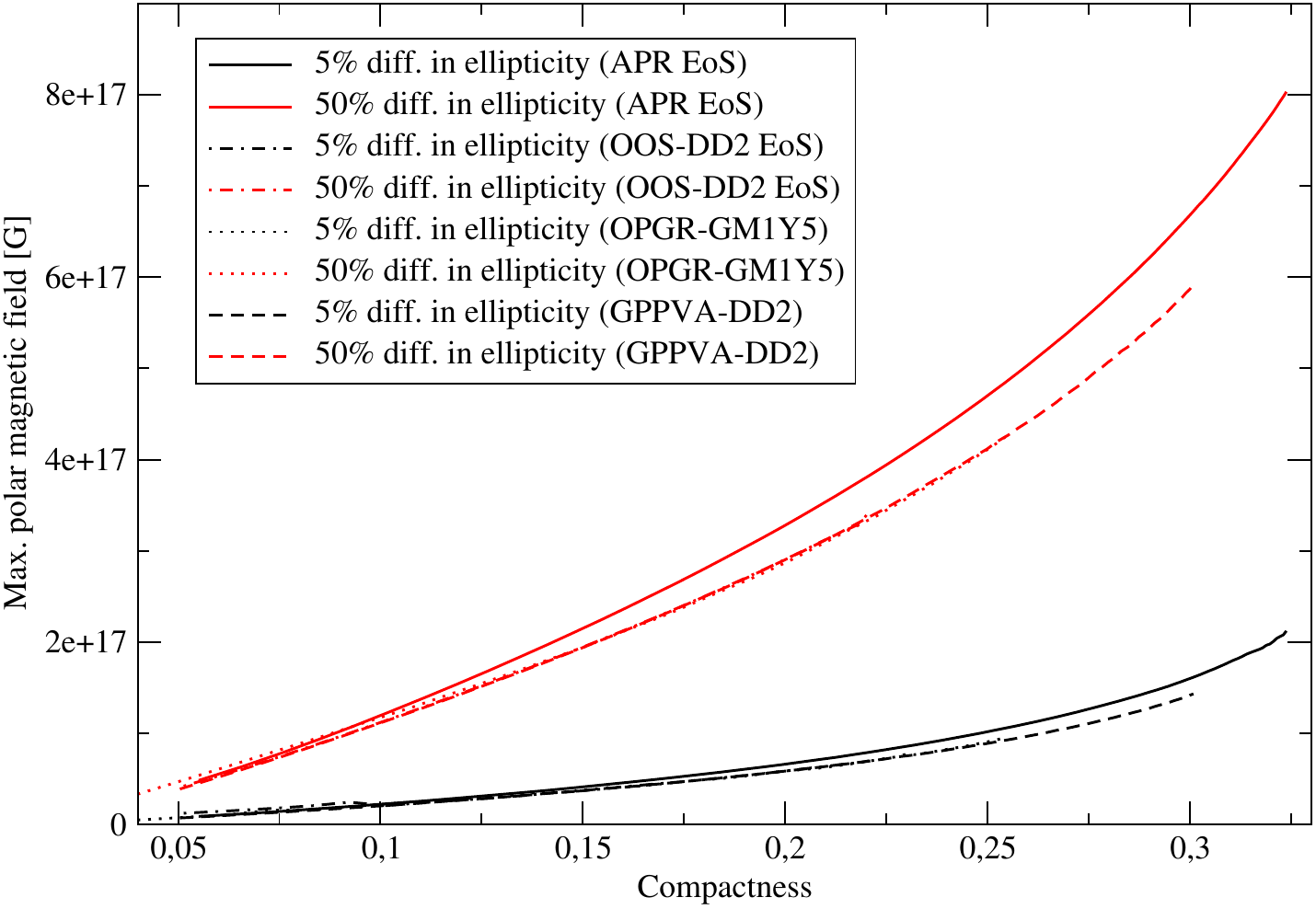}
  \caption{Maximal polar magnetic fields $B_{\rm pole}^5$ and $B_{\rm
      pole}^{50}$ for which the difference in surface ellipticity
    $\varepsilon_{\rm surf}$ between perturbative and fully numerical
    approaches are resp. lower than $5\%$ (black curves) and $50\%$
    (red curves), as functions of star's compactness for various
    EoSs. The curves end at the compactness corresponding to the EoS's
    maximum mass configuration. }
  \label{f:maxFields}
\end{figure}

In order to assess the dependence of the results on the EoS we have
used four different models. Although this is far from covering the
space of all possible EoSs, it gives an indication. These are the
DDFGOS(APR) model~\cite{davis-25}, the GPPVA(DD2)~\cite{grill-14}, the
OOS(DD2-FRG) model~\cite{otto-19,otto-20} and the
OPGR(GM1Y5)~\cite{oertel-15}. The three latter are based on covariant
density functionals to describe the hadronic interactions in dense
matter, the density dependent DD2 parameterization~\cite{typel-09} for
GPPVA(DD2) and OOS(DD2-FRG) and the nonlinear GM1 parameterization for
OPGR(GM1Y5). GPPVA(DD2) assumes a purely nucleonic composition,
whereas OOS(DD2-FRG) contains a phase transition to quark matter
modeled within a non-perturbative functional renormalization group
approach, here the version with 2+1 flavors and vector
interactions~\cite{otto-20} has been used. OPGR(GM1Y5) contains
hyperons in addition to nucleons. In the three cases, the nucleonic
part gives a rather stiff EoS, naturally softened by the onset of
hyperons or quarks above a certain density. The DDFGOS(APR) model is
based on the APR4 fit to variational calculations for the homogeneous
NS core~\cite{akmal-98}, extended in a unified way to the crust by
using the CUTER tool~\cite{davis-24,davis-25}. At high densities it
features a transition to a phase containing a pion condensate
resulting in a relatively soft EoS. All EoS are available in tabulated
form from the \textsc{CompOSE} database~\cite{compose, typel-22}, see
the entries \cite{ddfgapr,gppvaeos,oosdd2,gm1y5eos}.

The results presented in Secs.\ref{ss:mag_struct} and
\ref{ss:resu_deform} considered only the DDFGOS(APR) EoS. In
Fig.~\ref{f:maxFields}, the threshold polar magnetic field values
$B_{\rm pole}^5$ and $B_{\rm pole}^{50}$ are plotted as functions of
the compactness $\mathcal{C}$, for each of the four EoS discussed
above. For all EoS models, $B_{\rm pole}^5$ and $B_{\rm pole}^{50}$
are increasing with compactness, as expected since a more compact star
has a stronger gravitational field and thus requires a stronger
magnetic field to be deformed. Concerning the EoS dependence, the
three curves from CDF-based EoSs (DD2 and GM1) are very close one to
another and the only noticeable difference is the maximum compactness
reached, the APR one is slightly above. A possible explanation may
come from the fact that APR is a softer EoS, whereas the three others
are based on hadronic interaction models leading to stiffer behavior
up to the onset of possible additional particles at high
densities. This indicates that there is no ``universal'' behavior here
and that the threshold polar magnetic field values indeed depend on
the EoS. The influence of additional particles on the EoS at the
star's center at high compactnesses does not seem to play an important
role, but a more systematic study is on order to conclude on that
question.

\section{Summary and Conclusions}\label{s:conc}

We have compared, for the first time, two different techniques --
perturbative vs fully numerical -- to compute models of magnetized
neutron stars. The perturbative approach is based on a development
around a spherical background star and supposes first, that the
magnetic field is purely dipolar and second, that the deformation
induced by this field is small. The numerical technique relies on the
code \texttt{magstar}, which solves the coupled Einstein-Maxwell and
equilibrium equations with the assumption that the magnetic field is
purely poloidal. We have then studied for given values of the polar
magnetic field $B_{\rm pole}$, a quantity that is currently deduced
from pulsar timing observations, magnetic field intensity at the
star's center and equator, as well as quadrupole moment and surface
ellipticity. For each of these quantities, we have assessed the
interval for $B_{\rm pole}$ in which both approaches coincide up to
the accuracy used for the numerical integration.

On the one hand, for magnetic field values at the center and at the
equator with both approaches coincide up to
$B_{\rm pole} \sim 5\times 10^{16} - 10^{17}$~G, the precise value
depending on the star's compactness and on the EOS. Above this value,
higher order multipoles and effects from non-sphericity render the
perturbative approach inaccurate.

On the other hand, mass quadrupole moment and surface ellipticity
exhibit a somehow different behavior. For relatively low values of
$B_{\rm pole}$ they diverge, too, but this time it is the numerical
approach that fails to provide accurate enough results. This
presumably comes from the way these quantities are computed in the
code, which implies evaluating the difference between two numbers
having very close values. This automatically degrades the accuracy on
the result. For $B_{\rm pole} \lesssim 10^{15}$~G (depending again on
the star's compactness and EoS), the quadrupole is accurately computed
only with the perturbative approach, and similarly, below $B_{\rm
  pole} \lesssim 10^{10}$~G for the surface ellipticity. Then, at high
values of $B_{\rm pole}$ (higher than about $5\times 10^{16}$~G) the
perturbative approach fails for the computation of both $Q$ and
$\varepsilon_{\rm surf}$. Similar behaviors can be inferred if one
looks at the magnetic pressure and related quantities. The numbers
referenced above are taken from the Figs.~\ref{f:magfields} and
\ref{f:deform}, that were computed for the DDFGOS(APR)~EoS
of~\cite{davis-25} for a star with compactness
$\mathcal{C}=0.15$. These threshold values can be determined for other
EoSs and compactnesses using Fig.~\ref{f:maxFields}.

The magnetic polar field values above which the perturbative approach
ceases to be valid increase with the compactness of the star meaning
that, for higher compactness the perturbative approach is valid up to
higher values of $B_{\rm pole}$. Although there do not seem to be a
universal behavior and only a few EoSs have been studied, the
dependence on the EoS is weak. Moreover, a softer EoS seems to allow
the use of the perturbative approach up to higher values of the polar
magnetic field than a stiffer one. This feature seems to be linked to
the nucleonic part of the EoS and not to the possible presence of
additional particles (quarks or hyperons) at high densities, thus
higher compactness. All the points listed above and, in particular,
the values of the polar magnetic field below which the perturbative
approach gives accurate results in comparison with the full numerical
tool, are above the current maximal measured value of
SGR~1806-20~\cite{olausen-14}\footnote{This catalog is updated online
  \href{http://www.physics.mcgill.ca/~pulsar/magnetar/main.html}{http://www.physics.mcgill.ca/~pulsar/magnetar/main.html}}.
We conclude that \textsl{if rotation is neglected, the perturbative
  approach is valid for the modeling of all currently observed neutron
  stars}.

This result has some implications for the study of
magnetically-induced deformations that are considered as sources of
so-called continuous gravitational waves~\cite{bonazzola-96,
  mastrano-11}. Assuming that the current maximal measured value
represents an order-of-magnitude maximum for all pulsars and magnetars
potentially emitting gravitational waves, then it seems that a
perturbative approach should be sufficiently precise to compute their
deformations from an inferred polar magnetic field amplitude. In
the light of current searches for continuous gravitational waves from
the LIGO-Virgo-Kagra collaboration~\cite{lvk_pulsars-25, lvk_narrowband,
  lvk_allsky_isolated} this statement can make the analysis of a
possible future detection much easier, since it shows that a simpler
to code and computationally less expensive method could give similarly
accurate results than a fully numerical one.

One should however, keep in mind that this is the first study
comparing both types of approaches; it is therefore incomplete and
several paths for improvement are possible. A first improvement in the
numerical approach would certainly be a change in the way the
quadrupole moment and the surface ellipticity are computed, in order
to reach a better accuracy at low magnetic fields. Beside these
technical improvements, some better physical modeling is
needed. First, the comparison of mixed poloidal-toroidal magnetic
field configurations should be undertaken (see e.g.~\cite{das-23} for
polytropic EoSs).  The assumption of circularity in our
models prevents the existence of meridional currents and should be
relaxed if one wishes to study a more realistic magnetar setting with
toroidal currents and magnetic field. The perturbative approach with
such mixed magnetic fields has been described in
Ref.~\cite{colaiuda-08}, whereas the numerical approach has already
been devised by (at least) two groups: the XNS
code~\cite{pili-14}\footnote{\href{https://www.arcetri.inaf.it/science/ahead/XNS/index.html}{https://www.arcetri.inaf.it/science/ahead/XNS/index.html}},
and in Ref.~\cite{uryu-19} with the code \textsc{cocal}. Another
interesting line of research would be the inclusion of magnetic field
effects in the EoS, as well as the magnetization, following the model
presented in Ref.~\cite{chatterjee-15}. A more systematic study of the
EoS dependence and in particular the role of non-nucleonic degrees of
freedom at high densities should be undertaken, too.

Concerning the availability of our tools, the \texttt{magstar} code as
well as the EoS models are publicly available. For the future, we plan
to make the numerical tools for the integration of the perturbative
equations of Sec.~\ref{s:pert} publicly available, too.

\appendix
\section[\appendixname~\thesection]{}\label{s:app_dl}
Expanding the TOV Eqs.~(\ref{e:TOVr2z}) in Lindblom's form
\cite{lindblom-92} in log-enthalpy $h$ around the central value $h_c$ leads to
\begin{eqnarray}
r^2(h)\mkern-12mu&\xrightarrow[h\rightarrow h_c]{}&\mkern-12mu\frac{3(h_c-h)}{2\pi(3p_c+e_c)}\left(1+\frac{3\frac{de}{d h}|_{h_c}+15p_c-5e_c}{10(3p_c+e_c)}(h_c-h)\right)+O(h^3),\label{e:TOVinhIexp}\\
x(h)\mkern-12mu&\xrightarrow[h\rightarrow h_c]{}&\mkern-12mu\frac{2e_c(h_c-h)}{3p_c+e_c}\left(1-\frac{5e_c(e_c-3p_c)+3\frac{d e}{d h}|_{h_c}(6p_c+e_c) }{10e_c(3p_c+e_c)}(h_c-h)\right)+O(h^3) \label{e:TOVinhIIexp}.
\end{eqnarray}
This expansion can be used to deduce expansions around $r=0$ for the
metric potentials and thermodynamic quantities of the background solution
\begin{eqnarray}
\lambda(r)&\xrightarrow[r\rightarrow 0]{}&\frac{8}{3}\pi e_c r^2 + O(r^3), \label{e:tovr0I}\\
\nu(r)&\xrightarrow[r\rightarrow 0]{}&\nu_c+\frac{4}{3}\pi (3p_c+e_c) r^2 + O(r^3), \label{e:tovr0II} \\
p(r)&\xrightarrow[r\rightarrow 0]{}&p_c-\frac{2}{3}\pi (e_c+p_c)(3p_c+e_c) r^2 + O(r^3),\label{e:tovr0III} \\
e(r)&\xrightarrow[r\rightarrow 0]{}&e_c-\frac{2}{3}\pi (3p_c+e_c)\frac{d e}{d h}\Big|_{h_c} r^2 +O(r^3) \label{e:tovr0IV}.
\end{eqnarray}
The constant $\nu_c$ can be determined using the (exterior) Schwarzschild solution~(\ref{e:TOV_ext}) and the matching
condition~(\ref{e:nuDetla}) if $\nu(r)$ is required\footnote{The differential equations considered 
in this work require only $\nu'(r)$ and not $\nu(r)$ itself. Furthermore when working with the log-enthalpy $h$, 
Eq.~(\ref{e:nuhsol}) is available, which makes solving/integrating an ODE for $\nu'$ numerically not necessary.}.

Using Eqs.~(\ref{e:tovr0I})-(\ref{e:tovr0IV}) to expand Eq.~(\ref{e:Maxwell_l1}) in leading order in $r$ yields
\begin{eqnarray}
\frac{d^2a_1(r)}{dr^2} -a_1(r)+O(r^3)=0
\end{eqnarray}
which has the regular solution
\begin{eqnarray}
a_1(r)\mkern-12mu&\xrightarrow[r\rightarrow 0]{}&\mkern-12mu\alpha_0 r^2 = - \frac{1}{2}B_c r^2, \label{e:a1r0}
\end{eqnarray}
cf. Eq.~(\ref{e:B_part}).

Using the expansions~(\ref{e:tovr0I})-(\ref{e:tovr0IV}) and (\ref{e:a1r0}) we can then expand Eqs.~(\ref{e:h2}) and
(\ref{e:y2})
\begin{eqnarray}
  h_{2,H}(r)\mkern-12mu&\xrightarrow[r\rightarrow 0]{}&\mkern-12mu c_{h2H0} r^2 +O(r^3),\\
  y_{2,H}(r)\mkern-12mu&\xrightarrow[r\rightarrow 0]{}&\mkern-12mu-\frac{2}{3} c_{h2H0} \pi (3p_c+e_c)r^4+O(r^5),\\
  h_{2,P}(r)\mkern-12mu&\xrightarrow[r\rightarrow 0]{}&\mkern-12mu c_{h2P0} r^2 +O(r^3),\\
  y_{2,P}(r)\mkern-12mu&\xrightarrow[r\rightarrow 0]{}&\mkern-12mu -\frac{2}{3} \pi (e_c+3 p_c) c_{h2P0} r^4  -\frac{4}{3} \pi (e_c+p_c) \alpha_0 c_0 r^4 \notag\\
                       &&\qquad+\frac{16}{9} \pi (e_c+3 p_c) \alpha_0^2 r^4 +O(r^5),
\end{eqnarray}
where we set $a_1=j_1=0$ to obtain the homogeneous (H) results. Note
that defining $y_2$ (cf. Eq.~(\ref{e:def_y2})) instead of $k_2$ is
necessary to have a regular solution for $r\rightarrow 0$
\cite{colaiuda-08}. The constant of the particular
(P) solution $c_{h2P0}$ can be chosen as an arbitrary non-zero
value, while $c_{h2H0}$ and $K$ from Eqs.~(\ref{e:h2_vac}) and
(\ref{e:y2_vac}) have to be determined by matching interior and
exterior solutions for $h_2$ and $y_2$.

\section[\appendixname~\thesection]{}\label{s:app_match}
In the perturbative approach presented in Sec.~\ref{s:pert},
differential equations in the stellar exterior simplify due to the
vanishing of thermodynamic quantities, i.e. pressure, energy
density and currents. Analytic solutions in the exterior have already
been presented in the main part of this work. Matching (numerical)
interior and (analytical) exterior solutions has to be done by
imposing junction conditions at the stellar surface. We use a
variation of the \textsl{Israel-Darmois junction conditions} (IDJC),
see e.g. Ref.~\cite{raghoonundun-16} for a short overview and further
references. IDJC require continuity of the induced metric and induced
extrinsic curvature at the interface between the two spacetime
manifolds to be matched. Such a matching preserves spacetime
symmetries and lengths on the interface.

We will not go into details on how to formally derive conditions,
instead we refer the interested reader to work of B. Reina and R. Vera
\cite{reina-15}, which discusses matching of interior and exterior
solutions in Hartle-Thorne-like perturbative expansions such as the
one discussed here in detail. For the background star the matching
conditions are
\begin{eqnarray}
\nu^{\rm vac}(R)-\nu(R)&=&0,\label{e:nuDetla}\\
\nu'{}^{\rm vac}(R)-\nu'(R)&=&0,\\
\lambda^{\rm vac}(R)-\lambda(R)&=&0,\\
\lambda'{}^{\rm vac}(R)-\lambda'(R)&=&8\pi R e^{\lambda(R)} e(R).\label{e:dlambdaDetla}
\end{eqnarray}
The first discontinuity rises in the derivative of the $g_{rr}$
potential $\lambda'$ and it is proportional to the residual surface
density $e(R)$. For most EoSs $e(R)$ is very low, typically more than
ten orders of magnitude smaller than the central values. For all EoSs
discussed in this work $e(R)$ can thus be considered to vanish but for some
analytic interior models like the interior Schwarzschild solution or
exotic compact objects like pure quark stars $e(R)$ and the
resulting discontinuity in Eq.~(\ref{e:dlambdaDetla}) are
non-negligible.

For the magnetic field we employ magnetostatic matching conditions,
see e.g. Refs.~\cite{konno-99,rezzolla-04} for details.  The
components of the magnetic field tangential to the stellar surface are
continuous in the case of vanishing surface currents and the component
normal to the surface is always continuous. We will only consider
configurations with vanishing surface currents and therefore
completely continuous fields. Translating those matching conditions to
the vector potential using Eq.~(\ref{e:def_B}) necessitates continuity
of $a_1(r)$ and its first derivative at the stellar surface
\begin{eqnarray}
  a_1^{\rm vac}(R)&=&a_1(R),\\
  a_1'{}^{\rm vac}(R)&=&a_1'(R).
\end{eqnarray}

The metric potentials $h_2$ and $k_2$ related to the perturbative deformation induced by the magnetic field are continuous at the
stellar surface \cite{konno-99,reina-15} and thus by extension, considering the expressions discussed in this appendix also $y_2$
\begin{eqnarray}
  h_2^{\rm vac}(R)&=&h_2(R),\\
  k_2^{\rm vac}(R)&=&k_2(R),\\
  v_2^{\rm vac}(R)&=&v_2(R).
\end{eqnarray}
From the perturbations in $g_{rr}$, $m_2$ is continuous while $m_0$ is
not \cite{reina-15}. Similarly to $\lambda'{}$,
cf. Eq.~(\ref{e:dlambdaDetla}), the discontinuity in $m_0$ is
proportional to $e(R)$ and is important when computing the
perturbative shift/increase in gravitational mass due to magnetic
deformations.


\bibliography{biblio}

\end{document}